\documentclass[12pt]{article}

\usepackage{amssymb, amsthm, amsmath, amsfonts, eurosym, newtxmath, nccmath, cases}

\usepackage{graphicx, subfig}

\usepackage{booktabs, dcolumn, float, adjustbox}

\usepackage{hyperref}
\usepackage[font = small]{caption}

\usepackage[textsize = small]{todonotes}

\usepackage[backend = biber, style = apa, uniquename = false, sorting = ynt]{biblatex}

\usepackage{geometry, color, setspace, pdflscape, indentfirst}
\usepackage[bottom]{footmisc}

\newcolumntype{L}[1]{>{\raggedright\let\newline\\arraybackslash\hspace{0pt}}m{#1}}
\newcolumntype{C}[1]{>{\centering\let\newline\\arraybackslash\hspace{0pt}}m{#1}}
\newcolumntype{R}[1]{>{\raggedleft\let\newline\\arraybackslash\hspace{0pt}}m{#1}}

\definecolor{linkcolour}{rgb}{0, 0.2, 0.6}
\hypersetup{colorlinks, breaklinks, urlcolor = linkcolour, linkcolor = linkcolour, citecolor = linkcolour}

\newtheoremstyle{mystyle}
  {4pt} 
  {4pt} 
  {\itshape} 
  {0pt} 
  {\bfseries} 
  {.} 
  {.5em} 
  {} 
  
\theoremstyle{mystyle} 
\theoremstyle{mystyle} 
\theoremstyle{mystyle} \newtheorem{definition}{Definition}

\numberwithin{equation}{section}
\numberwithin{figure}{section}
\numberwithin{table}{section}
\numberwithin{assumption}{section}
\numberwithin{theorem}{section}

\usepackage[title]{appendix}

\usepackage[linesnumbered, ruled]{algorithm2e}

\usepackage{svg}

\DeclareMathOperator{\EX}{\mathbb{E}} 
\DeclareMathOperator{\Var}{\mathbb{V}} 
\DeclareMathOperator{\PX}{\mathbb{P}} 

\newcommand{\open}{“} 

\DeclareFontFamily{U}{mathx}{}
\DeclareFontShape{U}{mathx}{m}{n}{<-> mathx10}{}
\DeclareSymbolFont{mathx}{U}{mathx}{m}{n}
\DeclareMathAccent{\widehat}{0}{mathx}{"70}
\DeclareMathAccent{\widecheck}{0}{mathx}{"71}

\makeatletter
\let\save@mathaccent\mathaccent
\newcommand*\wideunderbar[1]{\@ifnextchar^{{\wide@underbar{#1}{0}}}{\wide@underbar{#1}{1}}}
\newcommand*\wide@underbar[2]{\if@single{#1}{\wide@underbar@{#1}{#2}{1}}{\wide@underbar@{#1}{#2}{2}}}
\newcommand*\wide@underbar@[3]{%
  \begingroup
  \def\mathaccent##1##2{%
    \let\mathaccent\save@mathaccent
    \if#32 \let\macc@nucleus\first@char \fi
    \setbox\z@\hbox{$\macc@style{\macc@nucleus}_{}$}%
    \setbox\tw@\hbox{$\macc@style{\macc@nucleus}{}_{}$}%
    \dimen@\wd\tw@
    \advance\dimen@-\wd\z@
    \divide\dimen@ 3
    \@tempdima\wd\tw@
    \advance\@tempdima-\scriptspace
    \divide\@tempdima 10
    \advance\dimen@-\@tempdima
    \ifdim\dimen@>\z@ \dimen@0pt\fi
    \rel@kern{0.6}\kern-\dimen@
    \if#31
      \underbar{\rel@kern{-0.6}\kern\dimen@\macc@nucleus\rel@kern{0.4}\kern\dimen@}%
      \advance\dimen@0.4\dimexpr\macc@kerna
      \let\final@kern#2%
      \ifdim\dimen@<\z@ \let\final@kern1\fi
      \if\final@kern1 \kern-\dimen@\fi
    \else
      \underbar{\rel@kern{-0.6}\kern\dimen@#1}%
    \fi
  }%
  \macc@depth\@ne
  \let\math@bgroup\@empty \let\math@egroup\macc@set@skewchar
  \mathsurround\z@ \frozen@everymath{\mathgroup\macc@group\relax}%
  \macc@set@skewchar\relax
  \let\mathaccentV\macc@nested@a
  \if#31
    \macc@nested@a\relax111{#1}%
  \else
    \def\gobble@till@marker##1\endmarker{}%
    \futurelet\first@char\gobble@till@marker#1\endmarker
    \ifcat\noexpand\first@char A\else
      \def\first@char{}%
    \fi
    \macc@nested@a\relax111{\first@char}%
  \fi
  \endgroup
}
\makeatother

\bibliography{references}

\begin{document}


\begin{titlepage}
    \title{Ordered Correlation Forest\thanks{I especially would like to thank Franco Peracchi for feedback and suggestions. I am also grateful to Matteo Iacopini, Michael Lechner, Jana Mareckova, Annalivia Polselli, seminar participants at University of Rome Tor Vergata and SEW-HSG research seminars, and conference participants at the WEEE 2023 for comments and discussions. Gabriel Okasa generously shared the code for implementing part of the DGPs in the simulation. The R package for implementing the methodology developed in this paper is available on CRAN at \href{https://cran.r-project.org/web/packages/ocf/index.html}{https://cran.r-project.org/web/packages/ocf/index.html}. The associated vignette is at \href{https://riccardo-df.github.io/ocf/}{https://riccardo-df.github.io/ocf/}.}}
    \author{Riccardo Di Francesco\thanks{Department of Economics and Finance, University of Rome Tor Vergata, Rome. Electronic correspondence: riccardo.di.francesco@uniroma2.it.}}
    
    \date{\today}
    
    \maketitle

    \vspace{-20pt}

    \begin{center}
        \href{https://riccardo-df.github.io/assets/papers/Ordered_Correlation_Forest.pdf}{Click here for the most recent version.}
    \end{center}
    
    \begin{abstract}
        \noindent Empirical studies in various social sciences often involve categorical outcomes with inherent ordering, such as self-evaluations of subjective well-being and self-assessments in health domains. While ordered choice models, such as the ordered logit and ordered probit, are popular tools for analyzing these outcomes, they may impose restrictive parametric and distributional assumptions. This paper introduces a novel estimator, the \textit{ordered correlation forest,} that can naturally handle non-linearities in the data and does not assume a specific error term distribution. The proposed estimator modifies a standard random forest splitting criterion to build a collection of forests, each estimating the conditional probability of a single class. Under an \open honesty" condition, predictions are consistent and asymptotically normal. The weights induced by each forest are used to obtain standard errors for the predicted probabilities and the covariates' marginal effects. Evidence from synthetic data shows that the proposed estimator features a superior prediction performance than alternative forest-based estimators and demonstrates its ability to construct valid confidence intervals for the covariates' marginal effects.
    
        \vspace{6pt}
   

        \bigskip

    \end{abstract}
    
    \setcounter{page}{0}
    
    \thispagestyle{empty}
\end{titlepage}

\pagebreak \newpage

\doublespacing


\section{Introduction}
\label{sec_introduction}
Categorical outcomes with a natural order, often referred to as ordered non-numeric outcomes, are commonly observed in empirical studies across the social sciences. For example, happiness research typically employs large surveys to collect self-evaluations of subjective well-being \parencite{frey2002can}, and health economics is heavily based on self-assessments in several health domains \parencite[see e.g.,][]{peracchi2012heterogeneity, peracchi2013heterogeneous}. These outcomes are usually measured on a discrete scale with five or ten classes, where the classes can be arranged in a natural order without any knowledge about their relative magnitude.

Ordered choice models, such as ordered logit and ordered probit, are frequently used to analyze the relationship between an ordered outcome and a set of covariates \parencite[see e.g.,][]{greene2010modeling}. These models target the estimation of the conditional choice probabilities, which represent the probability that the outcome belongs to a certain class given the values of the covariates. However, they are limited by their dependence on parametric and distributional assumptions that are often based on analytical convenience rather than knowledge about the underlying data generating process. As a result, econometricians may need to consider alternative techniques to produce more accurate and reliable predictions.

This paper introduces a novel machine learning estimator specifically optimized for handling ordered non-numeric outcomes. Employing traditional machine learning estimators \open off-the-shelf" can result in biased and inefficient estimation of conditional probabilities. This is because classification algorithms do not leverage the ordering information embedded in the structure of the outcome, and regression algorithms treat the outcome as if it is measured on a metric scale.\footnote{\ For comprehensive overviews of traditional classification and regression algorithms, the reader is referred to \textcite{hastie2009elements} and \textcite{efron_hastie_2016}.} The proposed estimator is designed to mitigate the biases that traditional methods can introduce, ultimately resulting in enhanced predictive performance.

The proposed estimator, named the \textit{ordered correlation forest}, adapts a standard random forest splitting criterion \parencite{breiman2001random} to the mean squared error relevant to the specific estimation problem at hand. The new splitting rule is then used to build a collection of forests, each estimating the conditional probability of a single class. After constructing the individual trees within each forest, the ordered correlation forest employs an unbiased estimator of conditional probabilities within each leaf. Model consistency is ensured, as the predictions always fall within the unit interval by construction. To estimate the covariates' marginal effects, the ordered correlation forest utilizes a nonparametric approximation of derivatives \parencite{lechner2019random}.

Under an \open honesty" condition \parencite{athey2016recursive}, the ordered correlation forest inherits the asymptotic properties of random forests, namely the consistency and asymptotic normality of their predictions \parencite{wager2018estimation}. Honesty is a subsample-splitting technique that requires that different observations are used to place the splits and compute leaf predictions and is crucial to achieving consistency of the random forest predictions.

The particular honesty implementation used by the ordered correlation forest allows for a weight-based estimation of the variance of the predicted probabilities. This is achieved by rewriting the random forest predictions as a weighted average of the outcomes \parencite{athey2019generalized}. The weights, which are obtained for the predicted probabilities, can be properly transformed to obtain standard errors for the covariates' marginal effects \parencite[for a similar approach, see][]{lechner2022modified, lechner2019random}. We can then use the estimated standard errors to conduct valid inference about the marginal effects as usual, e.g., by constructing conventional confidence intervals.

The rest of the paper unfolds as follows. Section \ref{sec_ordered_choice_models} provides a brief overview of ordered choice models and discusses some alternative estimation strategies. Section \ref{sec_estimation_and_inference} presents the ordered correlation forest, explaining estimation and inference about the statistical targets of interest. Section \ref{sec_simulation_results} uses synthetic data to compare the ordered correlation forest with alternative estimators and evaluate its performance in estimating and making inference about the covariates' marginal effects. Section \ref{sec_empirical_results} provides further comparisons with alternative estimators using real data. Section \ref{sec_conclusion} concludes. 


\section{Ordered Choice Models}
\label{sec_ordered_choice_models}
Ordered choice models are a class of statistical models used to analyze the relationship between an ordered non-numeric outcome $Y_i$ and a set of covariates $W_i$ \parencite{mccullagh1980regression}. These models are typically motivated by postulating the existence of a latent and continuous outcome variable of interest $Y_i^*$, assumed to obey the following regression model \parencite[see e.g.,][]{peracchi2014econometric}:

\begin{equation}
    Y_i^* = g \left( W_i \right) + \epsilon_i 
    \label{equation_regression_model_latent}
\end{equation}

\noindent where $W_i$ consists of a set of raw covariates, $g \left( \cdot \right)$ is a potentially non-linear regression function, and $\epsilon_i$ is independent of $W_i$ and has cumulative distribution $F \left( \cdot \right)$. Then, an observational rule links the observed outcome $Y_i$ to the latent outcome $Y_{i}^*$ using unknown threshold parameters $- \infty = \zeta_0 < \zeta_1 < \dots < \zeta_{M - 1} < \zeta_M = \infty$ that define intervals on the support of $Y_i^*$, with each interval corresponding to one of the $M$ categories or classes of $Y_i$:

\begin{equation}
    \begin{gathered}
        \zeta_{m - 1} < Y_i^* \leq \zeta_{m} \implies Y_i = m, \quad m = 1, \dots, M
    \end{gathered}
    \label{equation_observational_rule}
\end{equation}

\noindent Although the $M$ classes have a natural ordering, they are not measured on a cardinal scale. This limits our ability to make precise quantitative comparisons.

Researchers are typically interested in the estimation of the conditional choice probabilities, defined as:

\begin{equation}
    p_m \left( W_i \right) := \PX \left( Y_i = m | W_i \right)
\end{equation}

\noindent However, the marginal effect of the $j$-th covariate on $p_m \left( \cdot \right)$ is a more interpretable measure for ordered choice models. The marginal effect is defined differently depending on whether the $j$-th covariate is continuous or discrete:

\vspace{-0.9cm}

\begin{numcases}{\nabla^j p_m \left( w \right) := }
    \frac{\partial p_m \left( w \right)}{\partial w_j}, & \text{if } $w_j$ \text{ is continuous} \label{equation_marginal_effects_continuous} \\
    p_m \left( \lceil w_j \rceil \right) - p_m \left( \lfloor w_j \rfloor \right), & \text{if } $w_j$ \text{ is discrete} \label{equation_marginal_effects_discrete}
\end{numcases}

\noindent where $w_j$ is the $j$-th element of the vector $w$ and $\lceil w_j \rceil$ and $\lfloor w_j \rfloor$ correspond to $w$ with its $j$-th element rounded up and down to the closest integer. We can summarize the marginal effects in various ways, such as computing the marginal effect at the mean $\nabla^j p_m \left( \bar{w} \right)$, with $\bar{w}$ denoting a vector of means. Alternatively, we can compute the marginal effect at the median, the mean marginal effect, and the median marginal effect. 

From (\ref{equation_regression_model_latent}) and (\ref{equation_observational_rule}), the conditional choice probabilities write as:

\begin{equation}
    \begin{split}
        p_m \left( W_i \right) & = \PX \left( \zeta_{m - 1} < Y_i^* \leq \zeta_{m} | W_i \right) \\ 
        & = \PX \left( \zeta_{m - 1} - g \left( W_i \right) < \epsilon_i \leq \zeta_{m} - g \left( W_i \right) \right) \\ 
        & = F \left( \zeta_{m} - g \left( W_i \right) \right) - F \left( \zeta_{m - 1} - g \left( W_i \right) \right)
    \end{split}
    \label{equation_conditional_probabilities_expression}
\end{equation}

\noindent If the regression function $g \left( \cdot \right)$ and the distribution $F \left( \cdot \right)$ of the error term $\epsilon_i$ are known, we can estimate (\ref{equation_conditional_probabilities_expression}) directly using standard maximum likelihood methods.

However, in many practical applications, precise knowledge of $g \left( \cdot \right)$ is not available. Instead, a common approach is to approximate it using a linear-in-parameter model \parencite[see e.g.,][]{belloni2011high}:

\begin{equation}
    g \left( W_i \right) = X_i^T \beta + V_{i, k} 
    \label{equation_linear_approximation}
\end{equation}

\noindent where $X_i = h \left( W_i \right)$ is a $k$-dimensional vector of constructed covariates (generally the raw covariates $W_i$ plus interactions and polynomials thereof) and $V_{i, k}$ is an approximation error that is assumed to be independent of $X_i$. Substituting the linear approximation (\ref{equation_linear_approximation}) into (\ref{equation_regression_model_latent}) gives:

\begin{equation}
    \begin{gathered}
        Y_i^* = X_i^T \beta + U_i 
    \end{gathered}
\end{equation}

\noindent where the random error $U_i = \epsilon_i + V_{i, k}$ depends on $k$ through the approximation error $V_{i, k}$ and has cumulative distribution $G \left( \cdot \right)$. Then, we can approximate the statistical target $p_m \left( \cdot \right)$ as follows:\footnote{\ The ultimate target of estimation is $p_m \left( \cdot \right)$. $p_m^* \left( \cdot \right)$ serves as an approximation that allows us to tackle the estimation problem as if it were parametric.}

\begin{equation}
    \begin{split}
        p_m^* \left( W_i \right) & := \PX \left( Y_i = m | h \left( W_i \right) \right) \\ 
        & = G \left( \zeta_{m} - X_i^T \beta \right) - G \left( \zeta_{m - 1} - X_i^T \beta \right)
    \end{split}
    \label{equation_conditional_probabilities_expression_approximation}
\end{equation}

\noindent We can impose assumptions on the distribution $G \left( \cdot \right)$ of the random error $U_i$ to estimate (\ref{equation_conditional_probabilities_expression_approximation}) using standard maximum likelihood methods. Popular choices are the standard normal and the standard logistic distribution functions, producing the ordered probit and ordered logit models, respectively. In scenarios where $k > n$, regularization techniques such as L1- or L2-type penalization are needed.

Although easy to interpret and computationally simple, this approach features several limitations. First, it imposes strong distributional assumptions generally derived from analytical convenience rather than knowledge about the underlying data generating process. Second, it requires the specification of a linear-in-parameter model such as (\ref{equation_linear_approximation}) to account for non-linearities in $g(\cdot)$. Third, the estimated marginal effects have the restrictive property of single-crossing, meaning that they can change sign only once when moving from the smallest class to the largest.

Recent developments in statistical learning \parencite[see e.g.,][]{hastie2009elements, efron_hastie_2016} offer ways to overcome these limitations. For instance, random forest algorithms \parencite{breiman2001random} offer a nonparametric estimation approach that does not assume a specific error term distribution and can naturally handle non-linearities in $g \left( \cdot \right)$ without requiring a linear-in-parameter model. However, classification forests do not leverage the ordering information embedded in the structure of the outcome, and regression forests treat the outcome as if it is measured on a metric scale. Consequently, applying these algorithms \open off-the-shelf" can result in biased and inefficient estimation of conditional probabilities.

To overcome these limitations, one approach is to transform ordered non-numeric outcomes into a metric scale using scores based on the classes of the observed outcome, thus allowing us to use any regression algorithm on the transformed outcome. For example, \textcite{hothorn2006unbiased} propose using the midpoint values of the intervals defined on the support of the latent outcome as score values. In the cases where $Y_i^*$ is not observed, this translates into setting the scores equal to the class labels of $Y_i$. However, this assumes that the intervals are of equal length, which may not be accurate in practice. To address this issue, \textcite{hornung2020ordinal} proposes the ordinal forest estimator, which optimizes the class intervals and uses score values corresponding to these optimized intervals in a standard regression forest. The optimization process involves growing multiple forests using randomly generated candidate score sets, and constructing the final score values by summarizing the score sets with the smallest out-of-bag error. \textcite{hornung2020ordinal} shows that the ordinal forest estimator outperforms a standard regression forest that uses class labels as score values using both real and synthetic data. However, the optimization process can be computationally expensive, which may limit its practical use for large data sets or real-time applications.

Another approach involves expressing conditional probabilities as conditional expectations of binary variables, which can be estimated by any regression algorithm. One first strategy, which we label \textit{multinomial machine learning}, is to express conditional probabilities as follows:

\begin{equation}
    p_m \left( W_i \right) = \EX \left[ \vmathbb{1} \left( Y_i = m \right) | W_i \right] 
    \label{equation_multinomial_strategy}
\end{equation}

\noindent This allows us to estimate each $p_m \left( \cdot \right)$ separately by regressing the binary variable $\vmathbb{1} \left( Y_i = m \right)$ on $W_i$ using any nonparametric estimator:

\begin{equation}
    \hat{p}_m^{MML} \left( W_i \right) = \hat{p}_m \left( W_i \right)
\end{equation}

\noindent Alternatively, we can specify a linear-in-parameter model to estimate the approximate target $p_m^* \left( \cdot \right)$ through parametric regression of the binary variable $\vmathbb{1} \left( Y_i = m \right)$ on $X_i$. 

However, $\hat{p}_m^{MML} \left( \cdot \right)$ does not leverage the information embedded in the ordered structure of the outcome. To overcome this limitation, an alternative strategy that we label \textit{ordered machine learning} expresses conditional choice probabilities as the difference between the cumulative probabilities of two adjacent classes:

\begin{equation}
    \begin{split}
        p_m \left( W_i \right) & = \PX \left( Y_i \leq m | W_i \right) - \PX \left( Y_i \leq m - 1 | W_i \right) \\
        & = \mu_m \left( W_i \right) - \mu_{m - 1} \left( W_i \right)
    \end{split}
    \label{equation_ordered_strategy}
\end{equation}

\noindent with $\mu_m \left( W_i \right) := \EX \left[ \vmathbb{1} \left( Y_i \leq m \right) | W_i \right]$. Then we can estimate each $\mu_m \left( \cdot \right)$ separately by regressing the binary variable $\vmathbb{1} \left( Y_i \leq m \right)$ on $W_i$ using any nonparametric estimator and pick the difference between the cumulative probabilities of two adjacent classes to estimate $p_m \left( \cdot \right)$:\footnote{\ \textcite{lechner2019random} combine ordered machine learning with random forests \parencite{breiman2001random} and discuss how to estimate and conduct inference about marginal effects.}

\begin{equation}
    \hat{p}_m^{OML} \left( W_i \right) = \hat{\mu}_m \left( W_i \right) - \hat{\mu}_{m - 1} \left( W_i \right)
\end{equation}

\noindent As before, we can alternatively specify a linear-in-parameter model to estimate the approximate target $p_m^* \left( \cdot \right)$ through parametric regressions of the binary variables $\vmathbb{1} \left( Y_i \leq m \right)$ and $\vmathbb{1} \left( Y_i \leq m - 1 \right)$ on $X_i$.

However, $\hat{p}_m^{OML} \left( \cdot \right)$ can potentially produce negative predictions, thereby contradicting the definition of probabilities. Although we might resolve this issue by setting negative predictions to zero, such a solution is suboptimal, and an alternative estimator that does not require truncation may perform better. This paper introduces a novel estimator that leverages the ordered structure of the outcome and produces predictions that always fall within the unit interval, thus resulting in enhanced predictive performance compared to existing methods.


\section{Estimation and Inference}
\label{sec_estimation_and_inference}
In this section, I discuss the implementation of the ordered correlation forest (OCF) estimator. First, I illustrate the estimation of conditional choice probabilities and marginal effects. Second, I discuss the conditions required for the consistency and asymptotic normality of OCF predictions. Finally, I show how to conduct approximate inference about the statistical targets of interest.

\subsection{Estimation}
\label{sec_estimation}
Similar to the ordered machine learning approach, OCF computes the prediction of conditional choice probabilities as the difference between the cumulative probabilities of two adjacent classes (see equation \ref{equation_ordered_strategy}). However, instead of estimating $\mu_m \left( \cdot \right)$ and $\mu_{m - 1} \left( \cdot \right)$ separately, OCF internally performs this computation in a single random forest. This allows us to tie the estimation of $\mu_m \left( \cdot \right)$ and $\mu_{m - 1} \left( \cdot \right)$ to correlate the errors made in estimating these two expectations. Additionally, it avoids negative predictions.

To see the importance of correlating the estimation errors, we can decompose the mean squared error of a prediction $\hat{p}_m^{OML} \left( \cdot \right)$ at $w$ as follows:\footnote{\ This decomposition can be applied to any estimation strategy that involves calculating the difference between two surfaces. For example, \textcite{lechner2022modified} leverage this decomposition to estimate heterogeneous causal effects under a selection-on-observables assumption}.

\begin{equation}
    \begin{split}
            \text{MSE} \left( \hat{p}_m^{OML} \left( w \right) \right) & = \EX \left[ \left\{ \hat{p}_m^{OML} \left( w \right) - p_m \left( w \right) \right\}^2 \right] \\
            & = \EX \left[ \left\{ \hat{\mu}_m \left( w \right) - \hat{\mu}_{m - 1} \left( w \right) - \mu_m \left( w \right) + \mu_{m - 1} \left( w \right) \right\}^2 \right] \\
            & = \text{MSE} \left( \hat{\mu}_m \left( w \right) \right) + \text{MSE} \left( \hat{\mu}_{m - 1} \left( w \right) \right) - 2 \text{EC} \left( \hat{\mu}_m \left( w \right), \, \hat{\mu}_{m - 1} \left( w \right) \right)
    \end{split}
    \label{equation_mse_with_ec}
\end{equation}

\noindent where the last term is the error correlation and captures the degree to which the errors made in estimating $\mu_m \left( \cdot \right)$ and $\mu_{m - 1} \left( \cdot \right)$ are correlated:

\begin{equation}
    \text{EC} \left( \hat{\mu}_m \left( w \right), \, \hat{\mu}_{m - 1} \left( w \right) \right) = \EX\left[ \left\{ \hat{\mu}_m \left( w \right) - \mu_m \left( w \right) \right\} \left\{ \hat{\mu}_{m - 1} \left( w \right) - \mu_{m - 1} \left( w \right) \right\} \right]    
\end{equation}

\noindent Equation (\ref{equation_mse_with_ec}) shows that $\hat{p}_m^{OML} \left( \cdot \right)$ is a suboptimal estimator. Besides potentially leading to negative predictions, estimating $\mu_m \left( \cdot \right)$ and $\mu_{m - 1} \left( \cdot \right)$ separately minimizes only the mean squared error terms and ignores the error correlation. Tying the estimation of $\mu_m \left( \cdot \right)$ and $\mu_{m - 1} \left( \cdot \right)$ to correlate the errors could improve estimation performance since errors that move in the same direction cancel out when taking the difference $\hat{\mu}_m \left( \cdot \right) - \hat{\mu}_{m - 1} \left( \cdot \right)$. 

To address this limitation, OCF constructs a collection of forests, one for each of the $M$ classes of $Y_i$. However, rather than the standard criterion \parencite{breiman2001random}, OCF uses equation (\ref{equation_mse_with_ec}) as the splitting rule to build the individual trees in the $m$-th forest. This allows the estimator to account for the error correlation that $\hat{p}_m^{OML} \left( \cdot \right)$ ignores. Intuitively, during the tree-building process, OCF anticipates that the predictions in the final leaves will involve the difference between two estimated functions. Consequently, it seeks splits that not only yield accurate estimates of $\mu_m \left( \cdot \right)$ and $\mu_{m - 1} \left( \cdot \right)$ but also take into account the correlation between the errors made in estimating these expectations.

To use (\ref{equation_mse_with_ec}) as the splitting rule, we need to estimate its components. This, in turn, requires an estimator of $\mu_m \left( \cdot \right)$ in each node. An unbiased estimator of $\mu_m \left( \cdot \right)$ in a child node $\mathcal{C}_j \subset \mathcal{W}$ consists of the proportion of observations in $\mathcal{C}_j$ whose outcome is not greater than $m$:

\begin{equation}
    \check{\mu}_m \left( W_i \right) = \frac{1}{| \mathcal{C}_j |} \sum_{i : W_i \in \mathcal{C}_j} \vmathbb{1} \left( Y_i \leq m \right)
\end{equation}

\noindent This leads us to estimating $\text{MSE} \left( \check{\mu}_m \left( \cdot \right) \right)$ and $\text{EC} \left( \check{\mu}_m \left( \cdot \right), \check{\mu}_{m - 1} \left( \cdot \right) \right)$ in each node by their sample analogs:

\begin{flalign}
    \widehat{\text{MSE}}_j \left( \check{\mu}_m \left( W_i \right) \right) & = \frac{1}{| \mathcal{C}_j |} \sum_{i : W_i \in \mathcal{C}_j} \left[ \vmathbb{1} \left( Y_i \leq m \right) - \check{\mu}_m \left( W_i \right) \right]^2 \\ 
    \widehat{\text{EC}}_j \left( \check{\mu}_m \left( W_i \right), \check{\mu}_{m - 1} \left( W_i \right) \right) & = \frac{1}{| \mathcal{C}_j |} \sum_{i : W_i \in \mathcal{C}_j} \vmathbb{1} \left( Y_i \leq m \right) \vmathbb{1} \left( Y_i \leq m - 1 \right) - \check{\mu}_m \left( W_i \right) \check{\mu}_{m - 1} \left( W_i \right)
\end{flalign}

Then, in the $m$-th forest, OCF constructs individual trees by recursively partitioning each parent node $\mathcal{P} \subseteq \mathcal{W}$ into two child nodes $\mathcal{C}_1, \mathcal{C}_2 \subset \mathcal{P}$ such that the following minimization problem is solved:

\begin{equation}
    \min_{\mathcal{C}_1, \mathcal{C}_2} \, \sum_{j = 1}^2 \widehat{\text{MSE}}_j \left( \check{\mu}_m \left( W_i \right) \right) + \widehat{\text{MSE}}_j \left( \check{\mu}_{m - 1} \left( W_i \right) \right) - 2 \widehat{\text{EC}}_j \left( \check{\mu}_m \left( W_i \right), \check{\mu}_{m - 1} \left( W_i \right) \right)
\end{equation}

Once the recursive partitioning stops, each tree in the $m$-th forest unbiasedly estimates $p_m \left( \cdot \right)$ at $w$ by computing the proportion of observations in the same leaf as $w$ whose outcome equals $m$:

\begin{equation}
    \begin{split}
            \hat{p}_{m, b}^{OCF} \left( w \right) & = \check{\mu}_m \left( w \right) - \check{\mu}_{m - 1} \left( w \right) \\
            & = \frac{1}{| L_{m, b} \left( w \right) |} \sum_{i \in L_{m, b} \left( w \right) } \vmathbb{1} \left( Y_i = m \right)
    \end{split}
    \label{equation_ocf_trees_predictions}
\end{equation}

\noindent where $L_{m, b} \left( w \right)$ is the set of observations falling in the same leaf of the $b$-th tree as the prediction point $w$. The predictions from each tree are then averaged to obtain the forest predictions:\footnote{\ It may be necessary to perform a normalization step to ensure that $\sum_{m = 1}^M \hat{p}_m^{OCF} \left( w \right) = 1$. This is true also for $\hat{p}_m^{MML} \left( \cdot \right)$ and $\hat{p}_m^{OML} \left( \cdot \right)$.}

\begin{equation}
    \hat{p}_m^{OCF} \left( w \right) = \frac{1}{B_m} \sum_{b = 1}^{B_m} \hat{p}_{m, b}^{OCF} \left( w \right)
    \label{equation_ocf_conditional_probabilities}
\end{equation}

\noindent where $b = 1, \dots, B_m$ indexes the trees in the $m$-th forest. In contrast to ordered machine learning, OCF ensures model consistency, as the predictions $\hat{p}_m^{OCF} \left( \cdot \right)$ always fall within the unit interval by construction.

Estimation of marginal effects proceeds as proposed by \textcite{lechner2019random}. For discrete covariates, we can plug an estimate $\hat{p}_m^{OCF} \left( \cdot \right)$ of $p_m \left( \cdot \right)$ into equation (\ref{equation_marginal_effects_discrete}) to have a straightforward estimator of $\nabla^j p_m \left( \cdot \right)$:

\begin{equation}
    \nabla^j \hat{p}_m^{OCF} \left( w \right) = \hat{p}_m^{OCF} ( \lceil w_j \rceil ) - \hat{p}_m^{OCF} ( \lfloor w_j \rfloor )
    \label{equation_ocf_estimation_marginal_effects_discrete}
\end{equation}

\noindent For continuous covariates, we use a nonparametric approximation of the infinitesimal change in $w_j$:

\begin{equation}
    \nabla^j \hat{p}_m^{OCF} \left( w \right) = \frac{\hat{p}_m^{OCF} ( \widehat{\lceil w_j \rceil} ) - \hat{p}_m^{OCF} ( \widehat{\lfloor w_j \rfloor} )}{\widebar{w}_j - \wideunderbar{w}_j}
    \label{equation_ocf_estimation_marginal_effects_continuous}
\end{equation}

\noindent where $\widehat{\lceil w_j \rceil}$ and $\widehat{\lfloor w_j \rfloor}$ correspond to $w$ with its $j$-th element set to $\widebar{w}_j = w_j + \omega \, \sigma_j$ and $\wideunderbar{w}_j = w_j - \omega \, \sigma_j$, with $\sigma_j$ the standard deviation of $w_j$ and $\omega > 0$ a tuning parameter. 

\subsection{Asymptotic Properties}
\label{subsec_ocf_asymptotic}
\textcite{wager2018estimation} establish the consistency and asymptotic normality of random forest predictions. However, besides some regularity and technical assumptions, there are certain conditions regarding the construction of individual trees that must be satisfied. In the following, I discuss these conditions.

The first condition requires that the trees use different observations to place the splits and compute the leaf predictions. This condition is called \textit{honesty} and is crucial to bounding the bias of forest predictions.

\begin{definition}[\textit{Honesty}]
    A tree is honest if it uses the outcome $Y_i$ to either place the splits or compute the leaf predictions, but not both.
    \label{definition_honesty}
\end{definition}

\textcite{wager2018estimation} implement honesty by drawing a subsample $\mathcal{S}_b$ from the original sample $\mathcal{S}$ and splitting the subsample into two halves $\mathcal{S}_b^{tr}$ and $\mathcal{S}_b^{hon}$, using $\mathcal{S}_b^{tr}$ to grow the $b$-th tree and $\mathcal{S}_b^{hon}$ to compute its leaf predictions \parencite[see also][]{athey2019generalized}. Alternatively, \textcite{lechner2022modified} propose a different approach. They divide the original sample $\mathcal{S}$ into a training sample $\mathcal{S}^{tr}$ and an honest sample $\mathcal{S}^{hon}$, constructing trees from random subsamples of $\mathcal{S}^{tr}$ and computing their leaf predictions from $\mathcal{S}^{hon}$. This strategy ensures that, under i.i.d. sampling, the weights assigned to individual units in $\mathcal{S}^{hon}$ are independent of the outcomes of other units, thus allowing for weight-based inference about leaf predictions and their transformations, such as marginal effects. OCF adopts this strategy as well (details in Section \ref{subsec_ocf_inference}). However, this strategy is somewhat less efficient than the approach proposed by \textcite{wager2018estimation}. This is because, under the latter approach, each data point $w$ will participate in both $\mathcal{S}_b^{tr}$ and $\mathcal{S}_b^{hon}$ of some trees, thus achieving honesty while making more efficient use of the data. However, under this approach, each weight can depend on other units' outcomes,  which limits the usage of the weight-based representation of random forest predictions for obtaining standard errors for the leaf predictions and their transformations.

The second condition is that the leaves of the trees must become small in all dimensions of the covariate space as the sample size increases. This is necessary for achieving consistency of the predictions and is accomplished by introducing randomness in the tree-growing process and enforcing a regularity condition on how quickly the leaves get small.

\begin{definition}[\textit{Random-split}]
    A tree is random-split if, at every step of the tree-growing procedure, the probability that the next split occurs along the $j-$th covariate is bounded below by $\pi / k$, for some $0 < \pi \leq 1$, for all $j = 1, \dots, k$.
    \label{definition_random_split}
\end{definition}

\begin{definition}[\textit{$\alpha$-regularity}]
    A tree is $\alpha$-regular if each split leaves at least a fraction $\alpha$ of the observations in the parent node on each side of the split and the trees are fully grown to depth $d$ for some $d \in N$, that is, there are between $d$ and $2d - 1$ observations in each terminal node of the tree.
    \label{definition_alpha_regularity}
\end{definition}

To achieve $\alpha$-regularity, OCF ignores splits that do not satisfy this condition. The algorithm always selects the best split from among the candidate splits that would maintain at least a fraction $\alpha$ of the parent node's observations on both sides of the split. This way, we can rule out any influence of the splitting rule on the shape of the final leaves. 

Third, trees must be constructed using subsamples drawn without replacement, rather than bootstrap samples, as originally proposed by \textcite{breiman2001random}. 

Fourth, to establish asymptotic normality, trees must be symmetric.

\begin{definition}[\textit{Symmetry}]
    A predictor is symmetric if the (possibly randomized) output of the predictor does not depend on the order in which observations are indexed in the training and honest samples.
    \label{definition_symmetry}
\end{definition}

Under these conditions, \textcite{wager2018estimation} establish consistency and asymptotic normality of the random forest predictions. If the $M$ forests constructed by OCF satisfy these conditions, then they inherit these properties, thus producing consistent and asymptotically normally distributed predictions of conditional probabilities.

\subsection{Inference}
\label{subsec_ocf_inference}
In addition to the consistency and asymptotic normality of the random forest predictions, \textcite{wager2018estimation} show that the asymptotic variance of such predictions can be consistently estimated by adapting the infinitesimal jackknife estimator proposed by \textcite{wager2014confidence} to the case of subsampling without replacement. This approach can be used to estimate the variance of a prediction $\hat{p}_m^{OCF} \left( \cdot \right)$ at $w$. However, generalizing this method to estimate the variance of marginal effects $\nabla^j \hat{p}_m^{OCF} \left( \cdot \right)$ is not straightforward. 

To overcome this limitation, OCF employs an alternative approach that leverages the weight-based representation of random forest predictions \parencite{athey2019generalized} and adapts the weight-based inference proposed by \textcite{lechner2022modified} \parencite[see also][]{lechner2019random}. In particular, OCF implements honesty in a way that guarantees that the weight assigned to the $i$-th unit is independent of the outcomes of other units. This allows for the derivation of a straightforward formula for the variance of honest predicted probabilities and marginal effects.

First, we express OCF predictions as weighted averages of the outcomes. Let $\mathcal{S}$ denote the observed sample. The following provides an expression for a prediction $\hat{p}_m^{OCF} \left( \cdot \right)$ at $w$ numerically equivalent to that in (\ref{equation_ocf_conditional_probabilities}):

\begin{equation}
    \begin{gathered}
        \hat{p}_m^{OCF} \left( w \right) = \sum_{i \in \mathcal{S}} \hat{\alpha}_{m, i} \left( w \right) \vmathbb{1} \left( Y_i = m \right) \\
        \hat{\alpha}_{m, b, i} \left( w \right) = \frac{\vmathbb{1} \left( W_i \in L_{m, b} \left( w \right) \right)}{\left| L_{m, b} \left( w \right) \right|}, \quad \hat{\alpha}_{m, i} \left( w \right) = \frac{1}{B_m} \sum_{b = 1}^{B_m} \hat{\alpha}_{m, b, i} \left( w \right)
    \end{gathered}
    \label{equation_weight_based_probabilities}
\end{equation}

\noindent where the weights $\hat{\alpha}_{m, 1} \left( w \right), \dots, \hat{\alpha}_{m, |\mathcal{S}|} \left( w \right)$ determine the forest-based adaptive neighborhood of $w$. They represent how often the $i$-th observation in $\mathcal{S}$ shares a leaf with $w$ in the $m$-th forest. This measures how important the $i$-th observation is for fitting $p_m \left( \cdot \right)$ at $w$. Notice that $\sum_{i \in \mathcal{S}} \hat{\alpha}_{m, i} \left( w \right) = 1$ for all $w$.

Calculating the variance of a prediction $\hat{p}_m^{OCF} \left( w \right)$ in (\ref{equation_weight_based_probabilities}) is challenging because the weight assigned to the $i$-th unit $\hat{\alpha}_{m, i} \left( w \right)$ is a function of both $\mathcal{S}$ and $W_i$. Thus, this weight depends on the outcomes of all other units in $\mathcal{S}$, which complicates the formula for the variance.

However, the formula for the variance simplifies under the particular honesty implementation of OCF. Let $\mathcal{S}^{tr}$ and $\mathcal{S}^{hon}$ be a training sample and an honest sample obtained by randomly splitting the observed sample $\mathcal{S}$. Also, let $\hat{\alpha}_{m, i}^{tr} \left( \cdot \right)$ be the weights induced by a forest constructed using only $\mathcal{S}^{tr}$. Then, an honest prediction $\tilde{p}_m^{OCF} \left( \cdot \right)$ at $w$ is obtained by the following weighted average of observations in $\mathcal{S}^{hon}$:

\begin{equation}
    \tilde{p}_m^{OCF} \left( w \right) = \sum_{i \in \mathcal{S}^{hon}} \hat{\alpha}_{m, i}^{tr} \left( w \right) \vmathbb{1} \left( Y_i = m \right)
    \label{equation_weight_based_probabilities_honest}
\end{equation} 

\noindent The new weight assigned to the $i$-th unit $\hat{\alpha}_{m, i}^{tr} \left( w \right)$ is a function of $\mathcal{S}^{tr}$ and of $W_i$. Thus, under i.i.d. sampling this weight is independent of the outcomes of other units in $\mathcal{S}^{hon}$. This allows us to derive a simple formula for the variance of an honest prediction $\tilde{p}_m^{OCF} \left( w \right)$:

\begin{equation}
    \Var \left( \tilde{p}_m^{OCF} \left( w \right) \right) = | \mathcal{S}^{hon} | \Var \left( \hat{\alpha}_{m, i}^{tr} \left( w \right) \vmathbb{1} \left( Y_i = m \right) \right) 
    \label{equation_variance_probabilities}
\end{equation}

\noindent We can estimate this variance by its sample analog.

By plugging (\ref{equation_weight_based_probabilities_honest}) into (\ref{equation_ocf_estimation_marginal_effects_continuous}), we obtain the following estimator of honest marginal effects:\footnote{\ Similar results are obtained for discrete covariates by plugging  (\ref{equation_weight_based_probabilities_honest}) into (\ref{equation_ocf_estimation_marginal_effects_discrete}).}

\begin{equation}
    \begin{split}
        \nabla^j \tilde{p}_m^{OCF} \left( w \right) & = \frac{1}{\widebar{w}_j - \wideunderbar{w}_j} \left\{ \sum_{i \in \mathcal{S}^{hon}} \hat{\alpha}_{m, i}^{tr} ( \widehat{\lceil w_j \rceil} ) \vmathbb{1} \left( Y_i = m \right) - \sum_{i \in \mathcal{S}^{hon}} \hat{\alpha}_{m, i}^{tr} ( \widehat{\lfloor w_j \rfloor} ) \vmathbb{1} \left( Y_i = m \right) \right\} \\
        & = \frac{1}{\widebar{w}_j - \wideunderbar{w}_j} \sum_{i \in \mathcal{S}^{hon}} \check{\alpha}_{m, i}^{tr} ( \widehat{\lceil w_j \rceil}, \widehat{\lfloor w_j \rfloor} ) \vmathbb{1} \left( Y_i = m \right)      
    \end{split}
    \label{equation_weight_based_marginal_effects_honest}
\end{equation}

\noindent with $\check{\alpha}_{m, i}^{tr} ( \widehat{\lceil w_j \rceil}, \widehat{\lfloor w_j \rfloor} ) = \hat{\alpha}_{m,i}^{tr} ( \widehat{\lceil w_j \rceil} ) - \hat{\alpha}_{m, i}^{tr} ( \widehat{\lfloor w_j \rfloor} )$ a transformation of the original weights. Using the same argument as before, under i.i.d. sampling the weight assigned to the $i$-th unit $\check{\alpha}_{m, i}^{tr} ( \widehat{\lceil w_j \rceil}, \widehat{\lfloor w_j \rfloor} )$ is independent of the outcomes of other units in $\mathcal{S}^{hon}$. Thus the variance of an honest marginal effect $\nabla^j \tilde{p}_m^{OCF} \left( w \right)$ can be expressed as follows:

\begin{equation}
    \Var \left( \nabla^j \tilde{p}_m^{OCF} \left( w \right) \right) = \frac{| \mathcal{S}^{hon} |}{\left( \widebar{w}_j - \wideunderbar{w}_j \right)^2} \Var \left( \check{\alpha}_{m, i}^{\scriptstyle_{tr}} ( \widehat{\lceil w_j \rceil}, \widehat{\lfloor w_j \rfloor} ) \vmathbb{1} \left( Y_i = m \right) \right) 
    \label{equation_variance_marginal_effects}
\end{equation}

\noindent As before, we can estimate this variance by its sample analog.

Following the discussion of Section \ref{subsec_ocf_asymptotic}, the honest predicted probabilities in (\ref{equation_weight_based_probabilities_honest}) are consistent and asymptotically normal, provided that the weights $\hat{\alpha}_{m, i}^{tr} \left( \cdot \right)$ are induced by a forest composed of $\alpha$-regular with $\alpha \leq 0.2$ and symmetric random-split trees grown using subsampling without replacement. With these conditions met, we can use the estimated standard errors of honest predicted probabilities $\tilde{p}_m^{OCF} \left( \cdot \right)$ to conduct valid inference as usual, e.g., by constructing conventional confidence intervals.

Furthermore, under the same conditions the honest marginal effects in (\ref{equation_weight_based_marginal_effects_honest}) are a linear combination of normally distributed predictions, and thus have a normal distribution as well. Therefore, we can also construct conventional confidence intervals for honest marginal effects $\nabla^j \tilde{p}_m^{OCF} \left( \cdot \right)$ using their estimated standard errors.


\section{Simulation Results} 
\label{sec_simulation_results}
This section uses synthetic data to evaluate the performance of the ordered correlation forest (OCF) estimator. In the next subsection, I present the DGPs employed in the simulation. Then, I compare OCF with various alternative methods in terms of estimating conditional choice probabilities. Finally, I assess the ability of OCF in estimating and making inference about the covariates’ marginal effects.

\subsection{Data-Generating Processes}
\label{subsec_dgps}
Latent outcomes are generated as in (\ref{equation_regression_model_latent}), with $\epsilon_i \sim \textit{logistic} \left( 0, 1 \right)$. Six raw covariates are generated as $W_{i, 1}, W_{i, 3}, W_{i, 5} \sim \mathcal{N} \left( 0, 1 \right)$ and $W_{i, 2}, W_{i, 4}, W_{i, 6} \sim \textit{Bernoulli} \left( 0.4 \right)$. Covariates are independent of one another and of $\epsilon_i$. We consider $W_{i, 5}$ and $W_{i, 6}$ as \open noise" covariates, as they enter the DPGs below with null coefficients.
 
I consider three designs that differ in the regression function $g \left( \cdot \right)$:

\begin{flalign}
    \,\,\,\,\,\,\,\,\,\,\,\,\,\,\,\,\,\,\,\,\,\,\, \textit{Design 1.} & \,\,\, g \left( W_i \right) = W_i^T \beta \notag \\
    \,\,\,\,\,\,\,\,\,\,\,\,\,\,\,\,\,\,\,\,\,\,\, \textit{Design 2.} & \,\,\, g \left( W_i \right) = \sum_{j = 1}^{6} \sin \left( 2 W_{i, j} \right) \beta_j \notag \\
    \,\,\,\,\,\,\,\,\,\,\,\,\,\,\,\,\,\,\,\,\,\,\, \textit{Design 3.} & \,\,\, g \left( W_i \right) = 2 \sin \left( W_i^T \beta \right) \notag 
\end{flalign}

\noindent with $\beta = \left( 1, 1, 1/2, 1/2, 0, 0 \right)$ in all designs. \textit{Design 1} represents a linear model where all the raw covariates enter without transformation, serving as a benchmark for assessing the performance of the estimators under a straightforward and interpretable setting. In \textit{Design 2}, the covariates are transformed while preserving the additive structure of the model, thus allowing us to evaluate the estimators' ability to handle non-linearities arising from covariate transformations. \textit{Design 3} introduces more complex non-linearities by departing from the additive model structure and employing a nonlinear regression model. For each design, I consider four sample sizes, $| \mathcal{S} | \in \left\{ 500, \, 1000, 2000, 4000 \right\}$. Thus, I consider overall twelve different scenarios.

In each design, I obtain the observed outcomes $Y_i$ by discretizing $Y_i^*$ into three classes:

\begin{equation*}
    \zeta_{m - 1} < Y_i^* \leq \zeta_{m} \implies Y_i = m, \,\,\, m = 1, 2, 3
\end{equation*}

\noindent I construct the threshold parameters $\zeta_1$ and $\zeta_2$ as follows. First, I fix two values $\zeta_1^q = 0.33$ and $\zeta_2^q = 0.66$. Then, I generate a sample of $1,000,000$ $Y_i^*$ and set $\zeta_m = Q \left( \zeta_m^q \right)$, with $Q \left( \cdot \right)$ the empirical quantile function of $Y_i^*$. This way, the threshold parameters are uniformly spaced, and the class widths are approximately equal.

\subsection{Conditional Probabilities}
\label{subsec_conditional_probabilities}
After drawing a sample $\mathcal{S}$, I estimate the conditional choice probabilities using both multinomial and ordered machine learning techniques, combining them with random forests \parencite{breiman2001random} and penalized logistic regressions with an L1 penalty \parencite{tibshirani1996regression}. I refer to the resulting estimators as \textit{multinomial random forest} ($MRF$), \textit{multinomial L1 regression} ($ML1$), \textit{ordered random forest} ($ORF$), and \textit{ordered L1 regression} ($OL1$). I also consider two versions of OCF, the \open adaptive" version $OCF_{\scriptstyle_{A}}$ and the \open honest" version $OCF_{\scriptstyle_{H}}$. This way, we can quantify the loss in the precision derived from using fewer observations to build the forests, representing the price to pay for valid inference. Finally, I include the standard ordered logit ($LOGIT$) model as a parametric benchmark for comparison. 

To account for non-linearities in $g \left( \cdot \right)$, the parametric methods $LOGIT$, $ML1$, and $OL1$ employ different linear-in-parameter models such as (\ref{equation_linear_approximation}). Three different specifications are considered. The first specification consists of a model with only the raw covariates, that is, with $X_i = W_i$. The second specification introduces third-order polynomials for continuous covariates, leading to a set of $12$ covariates. The third specification enlarges this set by adding all the two-way interactions between the raw covariates, resulting in a total of $27$ covariates. In contrast, the forest-based estimators $MRF$, $ORF$, $OCF_{\scriptstyle_{A}}$, and $OCF_{\scriptstyle_{H}}$ are fed with only the raw covariates without adding any polynomials, interaction terms, or other transformations of the covariates, as these estimators can naturally handle non-linearities in $g \left( \cdot \right)$. To implement $OCF_{\scriptstyle_{H}}$, I randomly split $\mathcal{S}$ into a training sample $\mathcal{S}^{tr}$ used to construct the trees and an honest sample $\mathcal{S}^{hon}$ used to compute the leaf predictions. I choose $| \mathcal{S}^{tr} | = | \mathcal{S}^{hon} | = | \mathcal{S} |/2$. 

I rely on an external validation sample $\mathcal{S}^{val}$ of size $| \mathcal{S}^{val} | = 10,000$ to assess the predictive performance of the estimators. This large number of observations helps minimize the sampling variance. For each replication $r = 1, \dots, R$, I calculate the mean squared error, mean absolute error, and ranked probability score for each estimator:

\begin{flalign}
    \text{MSE}_r & = \frac{1}{| \mathcal{S}^{val} |} \sum_{i \in \mathcal{S}^{val}} \sum_{m = 1}^M  \left[ p_m \left( W_i \right) - \hat{p}_{m, r} \left( W_i \right) \right]^2 \label{equation_mean_squared_error} \\ 
    \text{MAE}_r & = \frac{1}{| \mathcal{S}^{val} |} \sum_{i \in \mathcal{S}^{val}} \sum_{m = 1}^M  \left| p_m \left( W_i \right) - \hat{p}_{m, r} \left( W_i \right) \right| \label{equation_mean_absolute_error} \\ 
    \text{RPS}_r & = \frac{1}{| \mathcal{S}^{val} |} \sum_{i \in \mathcal{S}^{val}} \frac{1}{M - 1} \sum_{m = 1}^M  \left[ \mu_m \left( W_i \right) - \hat{\mu}_{m, r} \left( W_i \right) \right]^2 \label{equation_ranked_score}
\end{flalign}

\noindent with $\hat{p}_{m, r} \left( \cdot \right)$ the estimated conditional probabilities in the $r$-th replication, and $\hat{\mu}_{m, r} \left( w \right) = \sum_{j = 1}^m \hat{p}_{j, r} \left( w \right)$ the estimated cumulative distribution function. Notice that, by simulation design, we can compute the true probabilities as in (\ref{equation_conditional_probabilities_expression}). I summarize these performance measures by averaging over the replications.\footnote{\ The objective of this subsection is to evaluate the prediction accuracy of each estimator. Thus, I do not consider the variance or the actual coverage rates of confidence intervals as performance measures, as these aspects are not relevant when the interest lies in prediction accuracy.}

\begingroup
  \setlength{\tabcolsep}{8pt}
  \renewcommand{\arraystretch}{1.1}
  \begin{table}[tbp]
    \centering
    \begin{adjustbox}{width = 1\textwidth}
    \begin{tabular}{@{\extracolsep{5pt}}l c c c c c c c c c c c c }
    \\[-1.8ex]\hline
    \hline \\[-1.8ex]
    & \multicolumn{4}{c}{\textit{Design 1}} & \multicolumn{4}{c}{\textit{Design 2}} & \multicolumn{4}{c}{\textit{Design 3}} \\ \cmidrule{2-5} \cmidrule{6-9} \cmidrule{10-13} 
     & 500 & 1,000 & 2,000 & 4,000 & 500 & 1,000 & 2,000 & 4,000 & 500 & 1,000 & 2,000 & 4,000 \\ 
    \addlinespace[2pt]
    \hline \\[-1.8ex] 

    \multicolumn{9}{l}{\textbf{\small Panel 1: $\overline{\text{MSE}}$}} \\
    $LOGIT_{raw}$ & 0.005 & 0.002 & 0.001 & 0.001 & 0.050 & 0.048 & 0.046 & 0.046 & 0.060 & 0.058 & 0.056 & 0.056 \\
    $LOGIT_{poly}$ & 0.009 & 0.004 & 0.002 & 0.001 & 0.029 & 0.025 & 0.022 & 0.021 & 0.051 & 0.046 & 0.043 & 0.042 \\
    $LOGIT_{int}$ & 0.020 & 0.010 & 0.005 & 0.002 & 0.041 & 0.030 & 0.025 & 0.023 & 0.035 & 0.023 & 0.018 & 0.016 \\
    $ML1_{raw}$ & 0.014 & 0.011 & 0.009 & 0.008 & 0.051 & 0.048 & 0.047 & 0.046 & 0.058 & 0.054 & 0.051 & 0.050 \\
    $ML1_{poly}$ & 0.015 & 0.010 & 0.007 & 0.006 & 0.033 & 0.027 & 0.024 & 0.022 & 0.051 & 0.045 & 0.042 & 0.040 \\
    $ML1_{int}$ & 0.016 & 0.009 & 0.005 & 0.003 & 0.040 & 0.031 & 0.026 & 0.024 & 0.038 & 0.026 & 0.020 & 0.016 \\
    $OL1_{raw}$ & 0.009 & 0.005 & 0.002 & 0.001 & 0.054 & 0.050 & 0.047 & 0.046 & 0.061 & 0.056 & 0.053 & 0.052 \\   
    $OL1_{poly}$ & 0.012 & 0.006 & 0.003 & 0.002 & 0.039 & 0.030 & 0.025 & 0.023 & 0.054 & 0.046 & 0.042 & 0.040 \\    
    $OL1_{int}$ & 0.016 & 0.008 & 0.004 & 0.002 & 0.048 & 0.036 & 0.029 & 0.025 & 0.048 & 0.032 & 0.023 & 0.018 \\ 
    $MRF$ & 0.045 & 0.035 & 0.028 & 0.022 & 0.046 & 0.037 & 0.029 & 0.022 & 0.055 & 0.043 & 0.034 & 0.026 \\
    $ORF$ & 0.050 & 0.044 & 0.040 & 0.036 & 0.054 & 0.048 & 0.043 & 0.040 & 0.061 & 0.050 & 0.042 & 0.037 \\
    $OCF_{\scriptstyle_{A}}$ & 0.044 & 0.038 & 0.035 & 0.032 & 0.046 & 0.040 & 0.037 & 0.034 & 0.054 & 0.044 & 0.037 & 0.033 \\
    $OCF_{\scriptstyle_{H}}$ & 0.035 & 0.025 & 0.018 & 0.014 & 0.040 & 0.030 & 0.022 & 0.017 & 0.054 & 0.041 & 0.030 & 0.022 \\ \cmidrule{1-13} 

    \multicolumn{9}{l}{\textbf{\small Panel 2: $\overline{\text{MAE}}$}} \\
    $LOGIT_{raw}$ & 0.093 & 0.065 & 0.045 & 0.032 & 0.296 & 0.289 & 0.286 & 0.284 & 0.310 & 0.304 & 0.301 & 0.300 \\
    $LOGIT_{poly}$ & 0.118 & 0.082 & 0.057 & 0.040 & 0.206 & 0.189 & 0.181 & 0.176 & 0.274 & 0.261 & 0.255 & 0.252 \\
    $LOGIT_{int}$ & 0.176 & 0.120 & 0.083 & 0.058 & 0.244 & 0.208 & 0.190 & 0.181 & 0.226 & 0.184 & 0.163 & 0.153 \\
    $ML1_{raw}$ & 0.153 & 0.131 & 0.118 & 0.111 & 0.306 & 0.296 & 0.291 & 0.289 & 0.326 & 0.313 & 0.307 & 0.303 \\
    $ML1_{poly}$ & 0.160 & 0.128 & 0.109 & 0.097 & 0.241 & 0.217 & 0.204 & 0.196 & 0.300 & 0.280 & 0.269 & 0.261 \\
    $ML1_{int}$ & 0.171 & 0.128 & 0.094 & 0.068 & 0.267 & 0.235 & 0.216 & 0.204 & 0.262 & 0.217 & 0.189 & 0.171 \\
    $OL1_{raw}$ & 0.127 & 0.090 & 0.063 & 0.045 & 0.317 & 0.301 & 0.292 & 0.288 & 0.327 & 0.310 & 0.302 & 0.298 \\   
    $OL1_{poly}$ & 0.145 & 0.102 & 0.073 & 0.052 & 0.261 & 0.225 & 0.204 & 0.191 & 0.306 & 0.278 & 0.263 & 0.253 \\    
    $OL1_{int}$ & 0.169 & 0.121 & 0.086 & 0.062 & 0.294 & 0.249 & 0.220 & 0.202 & 0.288 & 0.230 & 0.192 & 0.168 \\ 
    $MRF$ & 0.285 & 0.253 & 0.224 & 0.196 & 0.292 & 0.260 & 0.230 & 0.201 & 0.316 & 0.277 & 0.243 & 0.212 \\
    $ORF$ & 0.303 & 0.283 & 0.268 & 0.256 & 0.318 & 0.298 & 0.283 & 0.271 & 0.331 & 0.298 & 0.274 & 0.255 \\
    $OCF_{\scriptstyle_{A}}$ & 0.284 & 0.266 & 0.252 & 0.241 & 0.292 & 0.275 & 0.262 & 0.251 & 0.312 & 0.281 & 0.258 & 0.241 \\
    $OCF_{\scriptstyle_{H}}$ & 0.257 & 0.216 & 0.184 & 0.160 & 0.273 & 0.235 & 0.203 & 0.178 & 0.325 & 0.278 & 0.236 & 0.199 \\ \cmidrule{1-13} 

    \multicolumn{9}{l}{\textbf{\small Panel 3: $\overline{\text{RPS}}$}} \\
    $LOGIT_{raw}$ & 0.002 & 0.001 & 0.001 & 0.001 & 0.023 & 0.022 & 0.022 & 0.021 & 0.027 & 0.026 & 0.025 & 0.025 \\
    $LOGIT_{poly}$ & 0.004 & 0.002 & 0.001 & 0.001 & 0.013 & 0.011 & 0.010 & 0.009 & 0.022 & 0.020 & 0.019 & 0.019 \\
    $LOGIT_{int}$ & 0.008 & 0.004 & 0.002 & 0.001 & 0.018 & 0.013 & 0.011 & 0.010 & 0.014 & 0.010 & 0.007 & 0.006 \\
    $ML1_{raw}$ & 0.004 & 0.003 & 0.002 & 0.002 & 0.024 & 0.022 & 0.022 & 0.021 & 0.026 & 0.024 & 0.023 & 0.022 \\
    $ML1_{poly}$ & 0.005 & 0.003 & 0.002 & 0.002 & 0.014 & 0.012 & 0.010 & 0.010 & 0.022 & 0.019 & 0.018 & 0.017 \\
    $ML1_{int}$ & 0.006 & 0.003 & 0.002 & 0.001 & 0.018 & 0.014 & 0.012 & 0.010 & 0.015 & 0.010 & 0.007 & 0.006 \\
    $OL1_{raw}$ & 0.003 & 0.001 & 0.001 & 0.001 & 0.024 & 0.023 & 0.022 & 0.021 & 0.027 & 0.025 & 0.024 & 0.024 \\   
    $OL1_{poly}$ & 0.004 & 0.002 & 0.001 & 0.001 & 0.016 & 0.012 & 0.011 & 0.010 & 0.023 & 0.020 & 0.019 & 0.018 \\    
    $OL1_{int}$ & 0.005 & 0.003 & 0.001 & 0.001 & 0.019 & 0.015 & 0.012 & 0.011 & 0.017 & 0.011 & 0.008 & 0.006 \\  
    $MRF$ & 0.015 & 0.012 & 0.009 & 0.007 & 0.016 & 0.013 & 0.010 & 0.008 & 0.020 & 0.015 & 0.012 & 0.009 \\
    $ORF$ & 0.016 & 0.013 & 0.012 & 0.011 & 0.017 & 0.015 & 0.013 & 0.012 & 0.021 & 0.016 & 0.013 & 0.011 \\
    $OCF_{\scriptstyle_{A}}$ & 0.015 & 0.013 & 0.011 & 0.010 & 0.016 & 0.014 & 0.012 & 0.011 & 0.019 & 0.015 & 0.013 & 0.011 \\
    $OCF_{\scriptstyle_{H}}$ & 0.013 & 0.009 & 0.007 & 0.005 & 0.016 & 0.012 & 0.009 & 0.006 & 0.023 & 0.017 & 0.012 & 0.008 \\ 

    \addlinespace[-4pt]
    \\[-1.8ex]\hline
    \hline \\[-1.8ex]
    \end{tabular}
    \end{adjustbox}
    \caption{Comparison with alternative estimators. The three panels report the average over the replications of $\text{MSE}_r$ ($\overline{\text{MSE}}$), $\text{MAE}_r$ ($\overline{\text{MAE}}$), and $\text{RPS}_r$ ($\overline{\text{RPS}}$). The labels in the subscript of the parametric estimators $LOGIT$, $ML1$, and $OL1$ refer to the employed specification: $raw$ for only raw covariates, $poly$ for raw covariates plus third-order polynomials for continuous covariates, and $int$ for raw covariates plus third-order polynomials for continuous covariates plus all two-way interactions between the raw covariates.}
    \label{table_simulation_results}
  \end{table}
\endgroup

Table \ref{table_simulation_results} displays the results obtained with $R = 2,000$ replications. The simulation shows that OCF outperforms all other forest-based estimators uniformly across all considered scenarios. In particular, OCF consistently achieves lower MSE and MAE than $MRF$ and $ORF$, and minimal disparities in RPS are observed. An exception arises in \textit{Design 3} where, for the two smallest sample sizes, OCF and $MRF$ show similar performance.

The simulation also shows that OCF maintains a competitive performance when compared to the parametric estimators $LOGIT$, $ML1$, and $OL1$. As expected, when fed with only the raw covariates, $LOGIT$ and $OL1$ perform best in \textit{Design 1} since they correctly specify the parametric model and the distributional assumption of the error term. These estimators are closely followed by $ML1$ in terms of performance. However, the performance of $LOGIT$, $ML1$, and $OL1$ fed with only the raw covariates deteriorates when non-linearities in $g \left( \cdot \right)$ are introduced, causing them to rank among the worst estimators. For the smallest sample size, the performance gap with respect to OCF is relatively substantial in \textit{Design 2} (between $26$--$36\%$ in terms of MSE, between $8$--$16\%$ in terms of MAE, and between $43$--$48\%$ in terms of RPS) and moderate in \textit{Design 3} (around $11\%$ in terms of MSE and around $17\%$ in terms of RPS, with minimal disparities in MAE observed). However, in larger samples, this performance gap becomes more pronounced, with the MSE of $LOGIT$, $ML1$, and $OL1$ reaching values up to $165\%$ larger than that of OCF, MAE up to $61\%$, and RPS up to $232\%$.

Adding constructed covariates when the model for $g \left( \cdot \right)$ is linear and the raw covariates enter without transformation (\textit{Design 1}) deteriorates the performance of $LOGIT$ and $OL1$, as this primarily inflates the variance of the estimation. This effect becomes less relevant in larger samples. In contrast, when non-linearities in $g \left( \cdot \right)$ are introduced - either via transformations of the covariates (\textit{Design 2}) or by employing a non-linear regression model (\textit{Design 3}) - adding polynomials and interactions of the covariates significantly improves the performance of the parametric estimators. In particular, in \textit{Design 2}, $LOGIT$, $ML1$, and $OL1$ achieve their best performance by introducing third-order polynomials, with their performance deteriorating when interactions between covariates are also included, although this deterioration is substantially attenuated in larger samples. However, in \textit{Design 3}, where more complex non-linearities are introduced, including the interactions is necessary to achieve the best possible performance.

When constructed covariates are included in their specifications, $LOGIT$, $ML1$, and $OL1$ generally exhibit lower MSE, MAE, and RPS than OCF. However, in \textit{Design 2}, OCF outperforms all the parametric methods in larger samples, with advantages over the best parametric specification ranging between $22$--$31\%$ in terms of MSE, $7$--$10\%$ in terms of MAE, and $47$--$55\%$ in terms of RPS. An exception arises in the largest sample where $LOGIT$ and OCF tie in terms of MAE. Moreover, in \textit{Design 3}, the performance gap appears to narrow as the sample size increases, suggesting that OCF might outperform the parametric methods if enough observations are used to train the model.

Finally, we compare the adaptive and the honest versions of OCF to quantify the price to pay for valid inference. Surprisingly, the honest version $OCF_{\scriptstyle_{H}}$ performs better than the adaptive version $OCF_{\scriptstyle_{A}}$ in almost all scenarios despite using half of the observations to construct the forests. Honesty reduces the bias of the forests' estimates but generally comes at the expense of a higher variance. In this simulation, the reduction in bias appears to outweigh the increase in variance, resulting in improved prediction performance.

\subsection{Marginal Effects}
\label{sec_marginal_effects}
After drawing a sample $\mathcal{S}$, I split it into a training sample $\mathcal{S}^{tr}$ and an honest sample $\mathcal{S}^{hon}$ of equal size. Then, I use $\mathcal{S}^{tr}$ to construct the forests, and $\mathcal{S}^{hon}$ to estimate honest marginal effects at the mean and median of the covariates as in equation (\ref{equation_weight_based_marginal_effects_honest}). Additionally, I use the sample analog of equation (\ref{equation_variance_marginal_effects}) to get standard errors for the estimated effects.\footnote{\ Estimating the mean or the median marginal effect and its standard error would involve computing the weights $\check{\alpha}_{m, i}^{tr} \left( \cdot, \cdot \right)$ for each prediction point $w$, which would result in an impractically long computational time for a Monte Carlo exercise. Therefore, I restrict the analysis solely to the marginal effects at the mean and median of the covariates.}

To assess the performance of the estimator, I calculate the squared bias and variance for each marginal effect, as well as the actual coverage rates of their corresponding $95\%$ confidence intervals. Notice that, by simulation design, we can compute the true marginal effects. I summarize these performance measures by averaging across all marginal effects. 

Table \ref{table_simulation_results_me} displays the results obtained with $2,000$ replications. Overall, the simulation shows the ability of OCF to conduct asymptotically valid inference about marginal effects. The estimated squared bias consistently remains close to zero, indicating that the estimator is approximately unbiased. In smaller samples, the actual coverage rates of the confidence intervals tend to fall below the nominal rate and can be as low as $70\%$. However, as the sample size increases, the coverage rates gradually converge to the nominal level. In \textit{Design 2} and \textit{Design 3}, more observations are required to reach the nominal rate compared to \textit{Design 1}.

\begingroup
  \setlength{\tabcolsep}{8pt}
  \renewcommand{\arraystretch}{1.1}
  \begin{table}[b!]
     \centering
     \begin{adjustbox}{width = 1\textwidth}
     \begin{tabular}{@{\extracolsep{5pt}}l c c c c c c c c c c c c }
     \\[-1.8ex]\hline
     \hline \\[-1.8ex]
     & \multicolumn{4}{c}{\textit{Design 1}} & \multicolumn{4}{c}{\textit{Design 2}} & \multicolumn{4}{c}{\textit{Design 3}} \\ \cmidrule{2-5} \cmidrule{6-9} \cmidrule{10-13} 
      & 500 & 1,000 & 2,000 & 4,000 & 500 & 1,000 & 2,000 & 4,000 & 500 & 1,000 & 2,000 & 4,000 \\ 
     \addlinespace[2pt]
     \hline \\[-1.8ex] 

    \multicolumn{5}{l}{\textbf{\small Panel 1: Marginal effects at mean}} \\
    $\textit{Bias}^2$ & 0.002 & 0.001 & 0.001 & 0.001 & 0.010 & 0.009 & 0.010 & 0.011 & 0.011 & 0.009 & 0.009 & 0.010 \\
    $\textit{Var}$ & 0.010 & 0.012 & 0.014 & 0.017 & 0.012 & 0.015 & 0.018 & 0.022 & 0.011 & 0.013 & 0.016 & 0.019 \\
    $\textit{Coverage 95\%}$ & 0.84 & 0.90 & 0.93 & 0.95 & 0.80 & 0.85 & 0.90 & 0.92 & 0.71 & 0.79 & 0.86 & 0.91 \\ \cmidrule{1-13} 

    \multicolumn{5}{l}{\textbf{\small Panel 2: Marginal effects at median}} \\
    $\textit{Bias}^2$ & 0.002 & 0.001 & 0.001 & 0.001 & 0.006 & 0.003 & 0.001 & 0.001 & 0.013 & 0.009 & 0.005 & 0.002 \\
    $\textit{Var}$ & 0.010 & 0.012 & 0.014 & 0.017 & 0.012 & 0.015 & 0.018 & 0.022 & 0.011 & 0.013 & 0.016 & 0.019 \\
    $\textit{Coverage 95\%}$ & 0.85 & 0.90 & 0.94 & 0.95 & 0.81 & 0.88 & 0.93 & 0.95 & 0.70 & 0.76 & 0.83 & 0.89 \\ 

    \addlinespace[-4pt]
    \\[-1.8ex]\hline
    \hline \\[-1.8ex]
    \end{tabular}
    \end{adjustbox}
    \caption{Estimation and inference about the covariates' marginal effects. The first panel reports results for the marginal effects at the mean, and the second panel reports results for the marginal effects at the median.}
    \label{table_simulation_results_me}
  \end{table}
\endgroup

When we compare these results with those presented in Table \ref{table_simulation_results}, an interesting pattern emerges: the actual coverage rates of the confidence
intervals tend to be worse when the predictive performance of $OCF_{\scriptstyle_{H}}$ is lower. This pattern aligns with the fact that OCF estimates marginal effects by post-processing its conditional probability predictions (see equations \ref{equation_ocf_estimation_marginal_effects_discrete}--\ref{equation_ocf_estimation_marginal_effects_continuous}), meaning that the quality of conditional probability estimation directly impacts the accuracy of marginal effects estimation. As evident in Table \ref{table_simulation_results}, $OCF_{\scriptstyle_{H}}$ performs best in \textit{Design 1} and exhibits relatively lower performance in \textit{Design 3} compared to \textit{Design 2}. Consequently, for any given sample size, we observe a larger estimated squared bias in \textit{Design 2} and \textit{Design 3} relative to \textit{Design 1}, which explains why more observations are needed to reach the nominal rate in these designs. However, as the sample size increases, the predictive performance of OCF improves, and thus the estimated bias decays asymptotically. Therefore, in larger samples, the estimated confidence intervals are more likely to be centered around the true estimand, resulting in actual coverage rates that converge to the nominal level.


\section{Empirical Results} 
\label{sec_empirical_results}
This section uses real data to compare the predictive performance of the ordered correlation forest estimator with the same estimators of Section \ref{subsec_conditional_probabilities}.

\begingroup
    \setlength{\tabcolsep}{8pt}
    \renewcommand{\arraystretch}{1}
    \begin{table}[b]
        \centering
        \begin{adjustbox}{width = 1\textwidth}
        \begin{tabular}{@{\extracolsep{5pt}}l c c l c l c}
            \\[-1.8ex]\hline
            \hline \\[-1.8ex]
            \multicolumn{7}{c}{\textbf{Data Sets}} \\ \cmidrule{1-7}
            Data set & Sample Size & Outcome & \multicolumn{3}{c}{Class range} & N. Covariates \\

            \addlinespace[2pt]
            \hline \\[-1.8ex]  
            \textit{vlbw} & 218 & Apgar score & 1 (Life-threatening) & -- & 9 (Optimal) & 10 \\ 
            \textit{mammography} & 412 & Last mammography & 1 (Never) & -- & 3 (Over a year) & 5 \\
            \textit{support} & 798 & Functional disability & 1 (None) & -- & 5 (Fatal) & 15 \\
            \textit{nhanes} & 1,914 & Health status & 1 (Excellent) & -- & 5 (Poor) & 26 \\
            \textit{wines} & 4,893 & Quality & 1 (Moderate) & -- & 6 (High) & 11 \\
            
            \addlinespace[-4pt]
            \\[-1.8ex]\hline
            \hline \\[-1.8ex]

            \end{tabular}
        \end{adjustbox}
    \caption{Summary of data sets, sorted in increasing order of sample size.}
    \label{table_summary_data}
    \end{table}
\endgroup

I utilize the same data sets considered by \textcite{janitza2016random}, \textcite{hornung2020ordinal}, and \textcite{lechner2019random}. These data sets differ in terms of the number of covariates, observations, and classes of the observed outcome. Table \ref{table_summary_data} provides a summary of the data sets. For further details on the background of each data set, the reader is referred to \textcite{janitza2016random}.  

To assess the prediction accuracy of each estimator, I employ a ten-fold cross-validation procedure. Specifically, I randomly divide each data set into ten folds $\mathcal{S}^1, \dots, \mathcal{S}^{10}$ with roughly equal sizes. For each fold $f = 1, \dots, 10$, I fit all the estimators using the observations from all the other folds except for $\mathcal{S}^f$. Then, I calculate the same performance measures of Section \ref{subsec_conditional_probabilities} using the held-out $\mathcal{S}^f$:

\begin{flalign}
    \text{MSE}_f & = \frac{1}{| \mathcal{S}^{f} |} \sum_{i \in \mathcal{S}^{f}} \sum_{m = 1}^M  \left[ \vmathbb{1} \left( Y_i = m \right) - \hat{p}_{m, f} \left( W_i \right) \right]^2 \label{equation_mean_squared_error_real} \\ 
    \text{MAE}_f & = \frac{1}{| \mathcal{S}^{f} |} \sum_{i \in \mathcal{S}^{f}} \sum_{m = 1}^M  \left| \vmathbb{1} \left( Y_i = m \right) - \hat{p}_{m, f} \left( W_i \right) \right| \label{equation_mean_absolute_error_real} \\ 
    \text{RPS}_f & = \frac{1}{| \mathcal{S}^{f} |} \sum_{i \in \mathcal{S}^{f}} \frac{1}{M - 1} \sum_{m = 1}^M  \left[ \vmathbb{1} \left( Y_i \leq m \right) - \hat{\mu}_{m, f} \left( W_i \right) \right]^2 \label{equation_ranked_score_real}
\end{flalign}

\noindent with $\hat{p}_{m, f} \left( \cdot \right)$ the estimated conditional probabilities using all the other folds except for $\mathcal{S}^f$, and $\hat{\mu}_{m, f} \left( w \right) = \sum_{j = 1}^m \hat{p}_{j, f} \left( w \right)$ the estimated cumulative distribution function. Finally, I repeat this process ten times. This approach eliminates the dependence of the results on a particular training-validation sample split.

\begin{figure}[t!]
    \centering
    \includegraphics[scale=0.55]{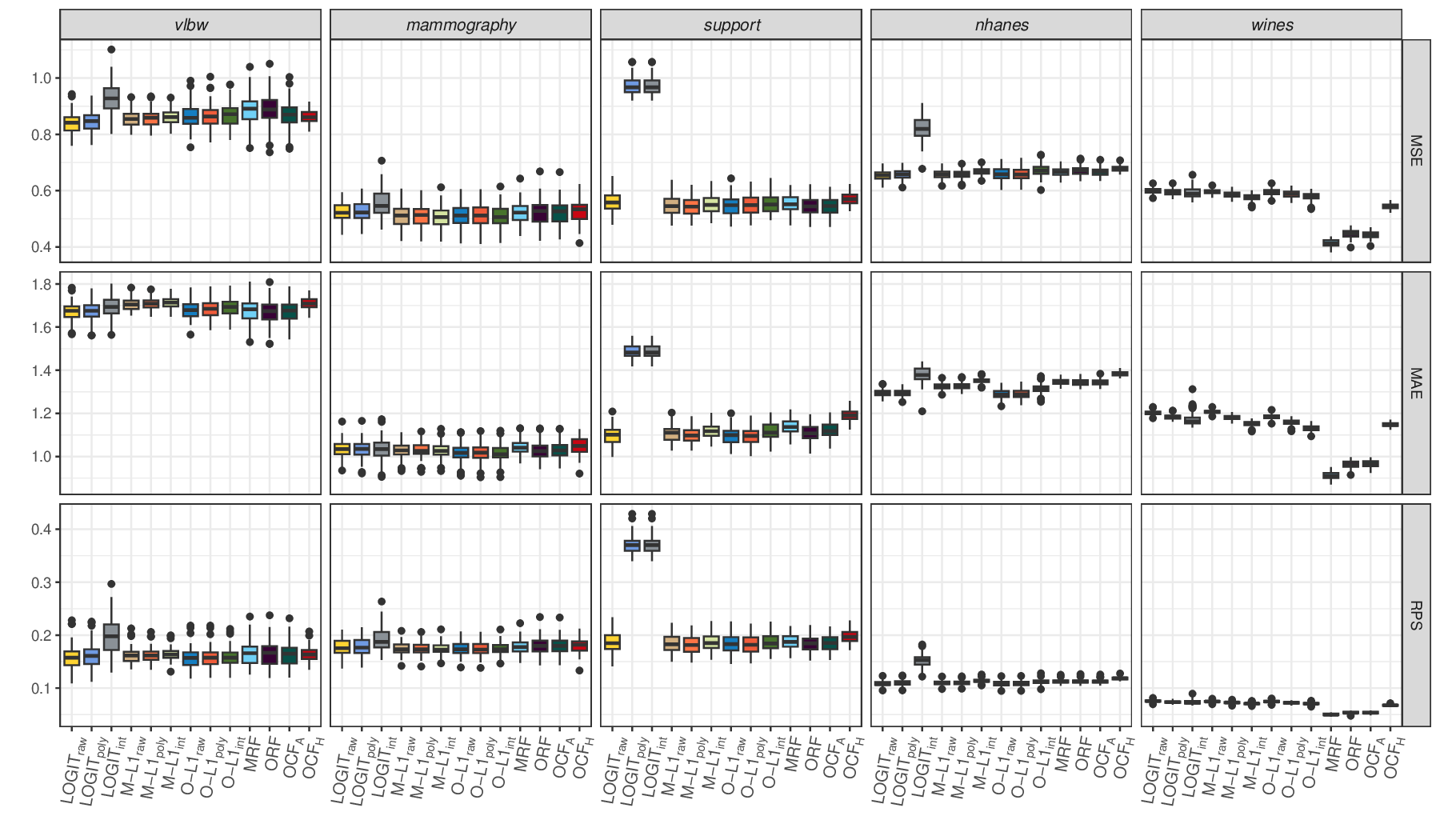}
    \caption{Prediction performance on real data sets. Each row contains boxplots showing the median and interquartile range of the estimated mean squared error (upper row), mean absolute error (mid row), and ranked probability score (lower row). Each column refers to a different data set, with the data set name displayed at the top of each column. Data sets are sorted according to their sample size.}
    \label{fig_empirical_results}
\end{figure}

Figure \ref{fig_empirical_results} reports the results by displaying boxplots showing the median and interquartile range of the estimated MSE, MAE, and RPS, together with their minima and maxima.\footnote{\ The cross-validation exercise yields a smaller sample size compared to the simulation results presented in Section \ref{subsec_conditional_probabilities}. Consequently, estimates of expected MSE, MAE, and RPS can be more imprecise and influenced by outliers. I report the distribution of the estimated MSE, MAE, and RPS using boxplots to provide a more robust assessment of the prediction performance of each estimator.}

Overall, the results indicate that OCF performs competitively compared to the other estimators, with no substantial differences in performance observed in most data sets. In the smallest data set under consideration (\textit{vlbw}), $LOGIT$, $ML1$, and $OL1$ perform marginally better than $MRF$, $ORF$, and OCF in terms of MSE, with similar MAE and RPS observed. However, in the largest data set (\textit{wines}), OCF emerges as one of the best estimators together with $MRF$ and $ORF$. This result highlights the advantage of forest-based methods over parametric methods in larger samples.

The addition of constructed covariates deteriorates the performance of $LOGIT$ while it does not substantially change the performance of $ML1$ and $OL1$, except in the largest data set (\textit{wines}) where it leads to improved performance for all parametric estimators. The deterioration in the performance of $LOGIT$ is particularly pronounced in the \textit{support} and \textit{nhanes} data sets. In these data sets, including third-order polynomials for continuous covariates and all the two-way interactions between the raw covariates results in a total of $324$ and $1394$ covariates, respectively. Given the moderate sample sizes, the inclusion of the additional covariates substantially increases the variance of the estimation, causing $LOGIT$ to perform worse compared to a specification with only the raw covariates. In contrast, the performance of $ML1$ and $OL1$ is not substantially affected by the inclusion of the additional covariates, as these estimators employ regularization techniques that help contain the increase in variance.

In contrast to the simulation results, the honest version of OCF does not outperform the adaptive version. In the two smaller data sets (\textit{vlbw} and \textit{mammography}), $OCF_{\scriptstyle_{A}}$ and $OCF_{\scriptstyle_{H}}$ exhibit similar MSE and RPS, with the MAE of $OCF_{\scriptstyle_{H}}$ being slightly larger than that of $OCF_{\scriptstyle_{A}}$. However, as the sample size increases, a performance gap in favor of $OCF_{\scriptstyle_{A}}$ emerges in terms of all performance measures. This gap becomes substantial in the largest data set.


\section{Conclusion} 
\label{sec_conclusion}

This paper proposes a novel machine learning estimator specifically optimized for handling ordered non-numeric outcomes. The proposed estimator adapts a standard random forest splitting criterion \parencite{breiman2001random} to the mean squared error relevant to the specific estimation problem at hand, thus mitigating the biases that traditional methods can introduce. The new splitting rule is then used to build a collection of forests, each estimating the conditional probability of a single class. A nonparametric approximation of derivatives is employed to estimate the covariates' marginal effects \parencite{lechner2019random}.

Under an \open honesty" condition \parencite{athey2016recursive}, the estimator inherits the asymptotic properties of random forests, namely the consistency and asymptotic normality of their predictions \parencite{wager2018estimation}. The particular honesty implementation used by the ordered correlation forest allows us to obtain standard errors for the covariates' marginal effects by leveraging the weight-based representation of the random forest predictions \parencite{athey2019generalized}. The estimated standard errors can then be used to construct asymptotically valid symmetric confidence intervals.

Evidence from synthetic data shows that the proposed estimator features a superior prediction performance than alternative forest-based estimators and demonstrates its ability to construct valid confidence intervals for the covariates' marginal effects. 

\newpage


\nocite{*}

\singlespacing
\newrefcontext[sorting = nty]
\newpage
\printbibliography
\newpage


\begin{appendices}

\doublespacing

\end{appendices}

\end{document}